\documentclass[epj]{svjour}
\usepackage{amsmath}
\usepackage{latexsym}
\usepackage{graphics}
\usepackage{rotating}
\usepackage{xspace}

\newcommand{\mettsm}{\mbox{\scriptsize ${\rm \not\! E}_{\rm T}$}}
\newcommand{\mettgmh}{\mbox{${\rm \not\! E}_{\rm T}>35$~GeV}}
\def \Et {{\rm E}_{\rm T}}
\def \Pt {{\rm P}_{\rm T}}
\newcommand{\met}{\mbox{${\rm \not\! E}_{\rm T}$}}
\newcommand{\Met}{\mbox{${\rm \not\! E}_{\rm T}$}} 
\newcommand{\metvec}{{\not\!\! \vec{E}_T}}

\def\eeggmet{ee\gamma\gamma\Met}
\def\gg{\gamma\gamma}
\def\lgmet{\ell\gamma\Met}
\def\llg{\ell\ell\gamma}
\def\egmet{e\gamma\Met}
\newcommand{\eg}{e\gamma}
\def\mgmet{\mu\gamma\Met}
\def\mugmet{\mu\gamma\Met}
\def\eeg{ee\gamma}

\def\mumug{\mu\mu\gamma}
\def\lg{\ell\gamma}
\def\lgg{\ell\gamma\gamma}
\newcommand{\lgX}{\ell\gamma\plus X}
\newcommand{\ggX}{\gamma\gamma\plus X}
\newcommand{\lplm}{\ell^+\ell^-}
\def\lum{{\cal L}}

\def\pbarp{{\bar p}p}
\def\ppbar{{\bar p}p}
\def\roots{{\sqrt s}}
\newcommand{\invpb}{pb^{-1}}

\newcommand{\invfb}{fb^{-1}}

\def\Z{Z^0}

\def\Wgg{W\gamma\gamma}
\def\Zgg{Z\gamma\gamma}
\def\Wg{W\gamma}
\def\Zg{Z\gamma}

\def\pizero{\pi ^0}
\def\epem{{\rm e^{+}e^{-}}}
\newcommand{\Zgstar}{Z^0\kern -0.25em/\kern -0.15em\gamma^*}
\newcommand{\goes}{\kern -0.18em\rightarrow\kern -0.18em}
\newcommand{\plus}{\kern -0.18em +\kern -0.18em}

\newcommand{\lsim}{\mbox{\small$\stackrel{<}{\sim}$\normalsize}}

\newcommand{\etal}{{\em et al.}}
\def\gt{>}

\newcommand{\lesssim } {\,\vcenter{\hbox{$\buildrel\textstyle<\over\sim$}}\,}

\newcommand{\ptt}{\mbox{$p_T$}}
\newcommand{\NONE}{\mbox{$\widetilde{\chi}_1^0$}}
\newcommand{\NTWO}{\mbox{$\widetilde{\chi}_2^0$}}
\newcommand{\CONE}{\mbox{$\widetilde{\chi}_1^{\pm}$}}

\newcommand{\none}{\NONE}
\newcommand{\ntwo}{\NTWO}
\newcommand{\cone}{\CONE}

\def\Gravitino{\tilde{G}}
\newcommand{\GeV}{\ensuremath{\mathrm{Ge\kern -0.1em V}}\xspace}
\newcommand{\Zgamma}{\Z\kern -0.1em/\kern -0.1em\gamma}
\newcommand{\GeVc}{\ensuremath{\mathrm{\ Ge\kern -0.1em V\kern -0.1em 
/\mathit{c}}}\xspace}
\newcommand{\GeVcsq}{\ensuremath{\mathrm{\ Ge\kern -0.1em V\kern -0.1em 
/\mathit{c}^2}}\xspace}

\begin{document}
\title{Searches for New Physics in Photon Final States}
\author{Andrey Loginov\inst{}
\thanks{\emph{For the CDF Collaboration}}%
}                     
\offprints{loginov@fnal.gov}          
\institute{ITEP, Moscow}
\date{Received: date / Revised version: date}
%
\abstract{
The Run I results on the searches for new physics in photon final
states were intriguing. The rare $\eeggmet$ candidate event and the
measured event rate for the signature $\ell+\gamma+\met$, which was
2.7 sigma above the Standard Model predictions, sparked
signature-based searches in the $\gamma\gamma+X$ and $\ell\gamma+X$
channels. With more data in Run II we should be able to answer a
simple question: was it an anomaly or were the Run I results the first
evidence for new physics? We present searches for New Physics in
Photon Final States at CDF Run II, Fermilab, with substantially more
data and a higher $\pbarp$ collision energy, 1.96~TeV, and the
upgraded CDF-II detector.
} 
\maketitle
\section{Introduction}
\label{introduction.section}
The Standard Model (SM)~\cite{SM} is an effective field theory that has so far
described the fundamental interactions of elementary
particles remarkably well. However the model breaks
down at energies of a few TeV, in that the cross-section for
scattering of longitudinal W bosons would otherwise violate unitarity.
The Fermilab Tevatron has the highest center-of-mass energy collisions
of any present accelerator, with $\roots =1.96$ TeV, and thus has the
potential to discover new physics.
As of September, 2005, the CDF experiment at Fermilab has recorded 1
$\invfb$ of data.  Physics results using 202 $\invpb$ to 345 $\invpb$
are presented in this paper.

\subsection{Motivation}
\label{motivation.section}
Why do we consider the photon final states a good signature for
observing new physics?

\begin{itemize}
\item Well Motivated Theories
   \begin{itemize}
     \item Most importantly Supersymmetry
   \end{itemize}
\item History
   \begin{itemize}
     \item Follow up on some of the anomalies from CDF in Run I~\cite{Toback_PRD,Toback_PRL,Ray_PRD,Jeff_PRD,Jeff_PRL}
   \end{itemize}
\item From the experimentalists' point of view, just because...
   \begin{itemize}
     \item  The photon is coupled to electric charge, and thus is radiated by all charged particles, including the incoming states (important for searching for invisible final states)
     \item The photon is massless and thus kinematically easier to produce than the W or Z
     \item The photon is stable, which implies a high acceptance, as there are no branching ratios to `pay'
     \item The photon is a boson and could be produced by a fermiphobic parent
     \item And if we then require
       \begin{itemize}
         \item Additional Lepton(s) $\Rightarrow$ high-$\Et$~\footnote{Transverse momentum and energy are defined as $\Pt =
p\sin\theta$ and $\Et = E\sin\theta$, respectively. The CDF coordinate system of $r$, $\varphi$, and $z$ is cylindrical,
with the $z$-axis along the proton beam.  The pseudorapidity is $\eta
= -\ln(\tan(\theta/2))$.} photon + high-$\Pt$ lepton + X signature is rare in SM, backgrounds are low for searches
         \item Additional Photon(s) $\Rightarrow$ the photons have moderate signal-to-noise but good efficiency and mass peak resolution
       \end{itemize}
   \end{itemize}
\end{itemize}

\subsection{Run I Results}
\label{runi.section}
\subsubsection{$\eeggmet$ Candidate Event}
\label{eeggmet.section}

\begin{figure}[!h]
\resizebox{0.5\textwidth}{!}{%
\includegraphics*{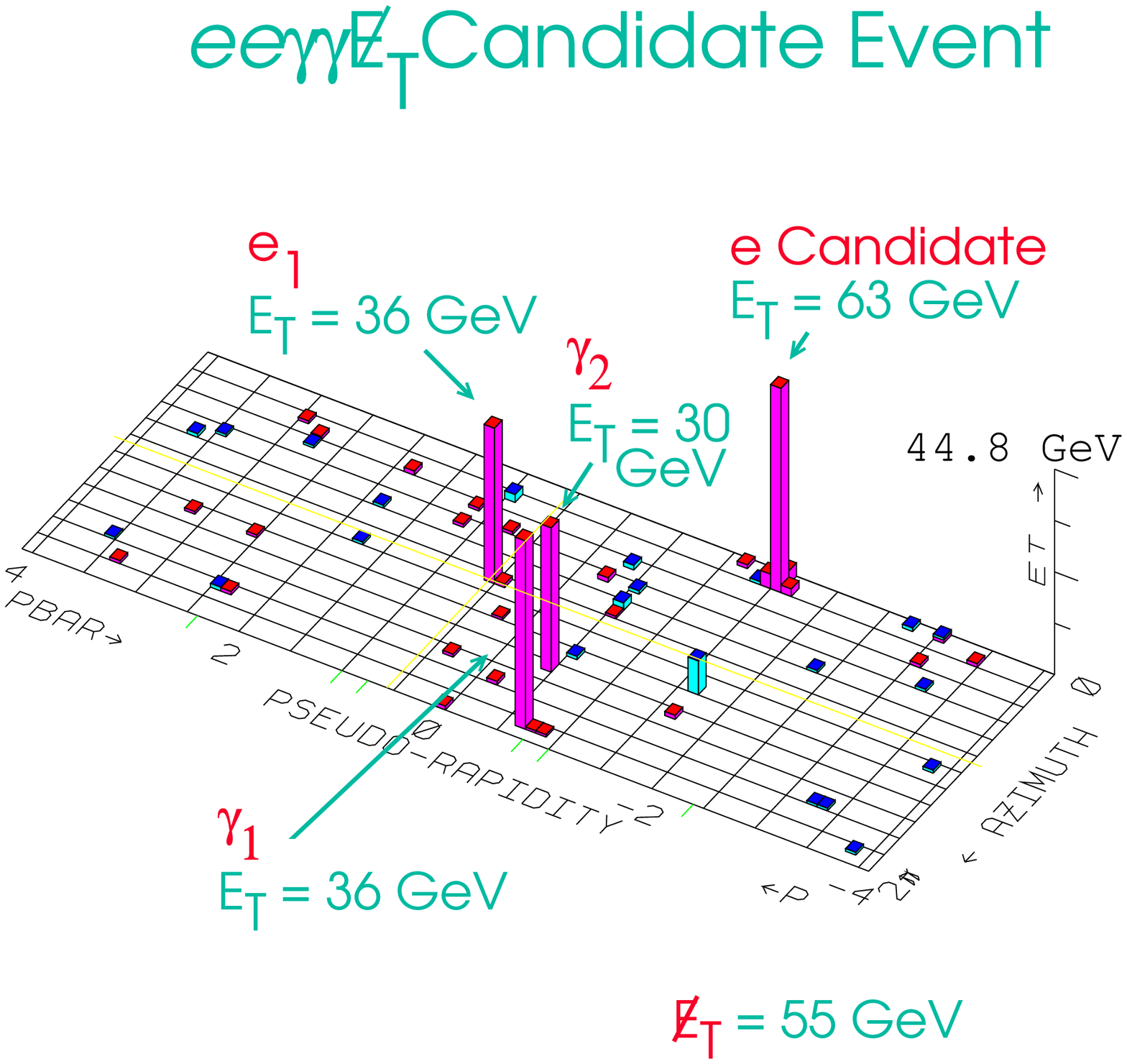}
}
\caption{The Run I $\eeggmet$ Candidate Event.}
\label{eeggmet.figure}       
\end{figure}

In 1995 the CDF experiment, measuring $\pbarp$ collisions at a
center-of-mass energy of 1.8 TeV at the Fermilab Tevatron, observed an
event~\cite{Toback_PRD,Toback_PRL,Toback_thesis} consistent with the
production of two energetic photons, two energetic electrons, and
large missing transverse energy~\footnote{Missing $\rm E_T$ ($\metvec$) is defined by $\metvec = -\sum_{i} E_T^i
\hat{n}_i$, where i is the calorimeter tower number for $|\eta| <$
3.6, and $\hat{n}_i$ is a unit vector perpendicular to the beam axis
and pointing at the i$^{th}$ calorimeter tower. We define the
magnitude $\met=|\metvec|$.}, $\met$
(Figure~\ref{eeggmet.figure}).

This signature is predicted to be very rare in the Standard Model of
particle physics, with the dominant contribution coming from the
WW$\gg$ production:

$WW\gamma\gamma\goes (e\nu)(e\nu)\gamma\gamma\goes
ee\gamma\gamma\met$,\\
from which we expect 8 $\times$ 10$^{-7}$
events. All other sources (mostly detector misidentification) lead to
5 $\times$ 10$^{-7}$ events. Therefore, we expect
(1 $\pm$ 1)$\times$ 10$^{-6}$ events, which would give us one
$\eeggmet$ candidate event if we had taken million times more data
than we actually had in Run I.

The event raised theoretical interest, however, as the two-lepton
two-photon signature is expected in some models of physics `beyond the
Standard Model' such as gauge-mediated models of
supersymmetry~\cite{susy}. For example, possible interpretation will
be:

$p\overline{p} \rightarrow {\tilde e}^+{\tilde e}^- (+ X)$,
${\tilde e} \rightarrow \ntwo + e$, $\ntwo \rightarrow \none\gamma$,\\
where ${\tilde e}$ is the selectron (the bosonic partner of the
electron), and $\none$ and $\ntwo$ are the lightest and
next-to-lightest neutralinos.

\subsubsection{$\gg$+X Search}
\label{ggx.section}
The detection of this single event led to the development of
`signature-based' inclusive searches to cast a wider net: in this case
one searches for two photons + X
~\cite{Toback_PRD,Toback_PRL,Toback_thesis}, where X stands for
anything, with the idea that if pairs of new particles were being
created these inclusive signatures would be sensitive to a range of
decay modes or to the creation and decay of different particle types.

In Run I Searches for {\bf$\gg$+X} all results were consistent with
the SM background expectations with no other exceptions
other than the observation of the $\eeggmet$ candidate
event(Table~\ref{ggx_runi.table})~\cite{Toback_PRL}.

\begin{table}[!t]
\begin{center}
\caption{Number of observed and expected  $\gamma\gamma$ events with additional
objects in 86 pb$^{-1}${\protect \cite{Toback_PRL}}.}
\label{ggx_runi.table}
\begin{tabular}{lcc}
\hline
Signature (Object) & Obs. & Expected \\
\hline
\mettgmh, $|\Delta\phi_{\mettsm-{\rm jet}}|>10^\circ$           & 1    & 0.5 $\pm$ 0.1       \\
N$_{\rm jet}\ge 4$, ${\rm E}_{\rm T}^{\rm jet}>10$~GeV, 
$|\eta^{\rm jet}|<2.0$                                          & 2    & 1.6 $\pm$ 0.4       \\
$b$-tag, ${\rm E}_{\rm T}^{b}>25$~GeV                           & 2    & 1.3 $\pm$ 0.7       \\
Central $\gamma$, ${\rm E}_{\rm T}^{\gamma_3}>25$~GeV           & 0    & 0.1 $\pm$ 0.1       \\
Central $e$ or $\mu$, ${\rm E}_{\rm T}^{e~{\rm or}~\mu}>25$~GeV & 3    & 0.3 $\pm$ 0.1       \\
Central $\tau$, ${\rm E}_{\rm T}^{\tau}>25$~GeV                 & 1    & 0.2 $\pm$ 0.1       \\
\hline
\end{tabular}
\end{center}
\end{table}

\subsubsection{From $\gg$ to $\lg$: $\lgX$ Search}
\label{lgx.section}
Another `signature-based' inclusive search, motivated by $\eeggmet$
event was for $\lgX$~\cite{Jeff_PRD,Jeff_PRL,Jeff_thesis}.

\begin{table}[!b]
\begin{center}
\caption{Run I Photon-Lepton Results: Number of observed and expected  $\ell\gamma$ events with additional
objects in 86 pb$^{-1}${\protect \cite{Jeff_PRL}}.}
\label{lgx_runi.table}
\begin{tabular}{l@{\extracolsep{0.0cm}}ccc}
\hline
Category & $\mu_{SM}$ & $N_0$ & P($N\ge N_0|\mu_{SM}$),\% \\
\hline
All $\lgX$           &          --           &           {\bf 77}     &   -- \\
\hline
Z-like $e\gamma$      &          --           &            17     &   -- \\
Two-Body   $\lg    X$ &      24.9$\pm$2.4  &            33     &   9.3 \\
Multi-Body $\lg    X$ &      20.2$\pm$1.7  &            27     &  10.0 \\
\hline
Multi-Body $\llg   X$ &      5.8  $\pm$ 0.6  &             5      &  68.0 \\
Multi-Body $\lgg   X$ &      0.02$\pm$0.02 &             1      &   1.5 \\
 Multi-Body $\lgmet X$ &
{\bf7.6  $\pm$ 0.7}        & 
{\bf16 }                   &
{\bf0.7} \\
\hline
\end{tabular}
\end{center}
\end{table}

In general data agrees with expectations, with the exception for the
$\lgmet$ category. We have observed 16 $\lgmet$ events on a background
of 7.6 $\pm$ 0.7 expected. The 16 $\lgmet$ events consist of 11
$\mugmet$ events and 5 $\egmet$ events, versus expectations of
4.2$\pm$0.5 and 3.4$\pm$0.3 events, respectively. The SM prediction
yields the observed rate of $\ell\gamma\met$ with {\bf 0.7\%}
probability (which is equivalent to {\bf 2.7} standard deviations for
a Gaussian distribution).

\begin{figure}[!h]
\centering
\resizebox{0.25\textwidth}{!}{%
\includegraphics*{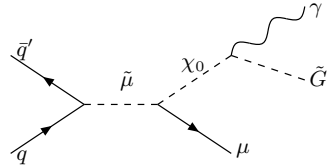}
}
\caption{Resonant smuon production and subsequent decay, producing the $\mu\gamma\met$ signature.}
\label{smuon_production.figure}
\end{figure}

One of the first SUSY interpretation of the CDF { $\mgmet$}
events~\cite{smuon_PRL} was resonant smuon $\tilde{\mu}$ production
with a single dominant R-parity violating
coupling (Figure~\ref{smuon_production.figure}).

The Run I search was initiated by an anomaly in the data itself, and
as such the 2.7 sigma excess above the SM expectations must be viewed
taking into account the number of such channels a fluctuation could
have occurred in.

\section{From Run I to Run II}
\label{runi_runii.section}

Having many different hints from the signature-based sear\-ches for new
physics in photon final states in Run I, the strategy for Run II was
straightforward: take more data. The main points were:

\begin{itemize}
\item Increase the Collision Energy: 1.80 $\rightarrow$ 1.96 TeV
\item Increase the rate at which we take data: 3500 $\rightarrow$ 396 ns (timing between bunches)
\item Upgrade the Detectors
\end{itemize}

\subsection{CDF Run II Detector}
\label{cdf_detector.section}

The CDF-II detector~\cite{CDFII} is a cylindrically symmetric
spectrometer designed to study $\pbarp$ collisions at the Fermilab
Tevatron, that uses the same solenoidal magnet and central
calorimeters as the CDF-I detector~\cite{CDFI} from which it was
upgraded. Because the analyses described here have been motivated by
the Run I searches, we note especially the differences from the Run I
detector relevant to the detection of photons, leptons, and $\met$.

The central calorimeters are physically unchanged; however, the readout
electronics has been replaced to accommodate the smaller proton and
anti-proton bunch spacing of the Tevatron in Run II.  The end-cap
(plug) and forward calorimeters have been replaced with a more compact
scintillator-based design, retaining the projective
geometry~\cite{cal_upgrade}.

The tracking system used to measure the momenta of charged particles
has been replaced, with the central outer tracker upgraded to have
smaller drift cells~\cite{COT}, and the inner tracking chamber and
silicon system replaced by a system of silicon strip chambers with
more layers, now in 2-dimensions~\cite{SVX}. The new inner tracking
system has substantially more material, resulting in more
bremsstrahlung (photons) produced by high-$\Pt$ electrons.

The central CMU, CMP, and CMX muon systems\footnote{The CMU (Central
Muon Chambers) system consists of gas proportional chambers in the
region $|\eta|<0.6$; the CMP (Central Muon Upgrade) system consists of
chambers after an additional meter of steel, also for
$|\eta|<0.6$. The CMX (Central Muon Extension) chambers cover
$0.6<|\eta|<1.0$.} are also physically unchanged in design, but the
coverage of the CMP and CMX muon systems~\cite{muon_systems} has been
extended by filling in gaps in $\varphi$~\cite{CDFII}.

\section{Run II: Searches for New Physics in Photon Final States}
\label{runii_searches.section}

The Run I results on the searches for new physics in photon final
states were
intriguing~\cite{Toback_PRD,Toback_PRL,Jeff_PRD,Jeff_PRL}. The rare
$\eeggmet$ candidate event and the measured event rate for the
signature $\ell+\gamma+\met$, which was 2.7 sigma above the SM
predictions, sparked signature-based searches in the $\gamma\gamma+X$
and $\ell\gamma+X$ channels.

With more data in Run II we should be able to answer a simple
question: was it an anomaly or were the Run I results the first
evidence for the new physics?

There are lots of searches involving photon final states at CDF in Run
II. Some of the analyses are presented in this paper:

\begin{itemize}
\item Search for High-Mass Diphoton State and Limits on Randall-Sundrum Gravitons              (Section~\ref{gg_runii.section})
\item Search for Anomalous Production of Diphoton Events with $\met$ and Limits on GMSB Models  (Section~\ref{ggmet_runii.section})
\item Search for Lepton-Photon-X Events                                                         (Section~\ref{lgx_runii.section})
\end{itemize}

\subsection{Search for High-Mass Diphoton State and Limits on Randall-Sundrum Gravitons             }
\label{gg_runii.section}

Searches for new particles decaying into two identical particles are
broad, inclusive and sensitive. The production of the new particle may
be direct or in association with other particles, or in a decay
chain. The discovery of a sharp mass peak over background would be a
compelling evidence for the production of a new particle. The diphoton
final state is important because the photons are bosons and the parent
may be fermiphobic. The photons have moderate signal-to-noise but good
efficiency and mass peak resolution.

One model producing a diphoton mass peak is Randall-Sundrum
gravitons~\cite{randall_sundrum}. Current string theory proposes that
as many as seven new dimensions may exist and the geometry of these
extra dimensions is responsible for gravity being so weak. The
Randall-Sundrum model has the property that a parameter, the warp
factor, determines the curvature of the extra dimensions and therefore
the mass of the Kaluza-Klein graviton resonances, which decay to two
bodies including photons.

Details on this analysis are reported in~\cite{www_cdf_gg}.

\subsubsection{Data Sample}
\label{gg_runii_data.section}

The sample corresponds to 345 $\invpb$ of data taken between February
2002 and July, 2004. We require that the data were taken under good
detector conditions for a reliable photon identification. We apply
selection cuts as follows:

\begin{itemize}
\item Photons in Central Calorimeter
\item $\Et^{\gamma}\gt$ 15 GeV
\item M ($\gamma, \gamma$) $\gt$ 30 GeV
\end{itemize}

To select a photon in a central calorimeter (approximately
$0.05<|\eta|<1.0$), we require a central electromagnetic cluster that: 
a)~is not near the boundary in $\phi$ of a calorimeter tower~\footnote{
The fiducial region has $\sim$87\% coverage in the central region.}
b)~have the ratio of hadronic to electromagnetic energy, 
Had/EM, $< 0.055+0.00045\times E^{\gamma}(\GeV)$; 
c)~have no tracks, or only one track with \mbox{$\ptt <1$~GeV/$c$},
extrapolating to the towers of the cluster; 
d)~is isolated in the calorimeter and tracking chamber~\footnote{
To reject hadronic backgrounds that fake prompt photons, candidates
are required to be isolated in the calorimeter and tracking
chamber. In the calorimeter the isolation is defined as the energy in
a cone of 0.4 in $\eta-\phi$ space, minus the photon cluster energy,
and corrected for energy loss into cracks as well as the number of
reconstructed $\ppbar$ interactions in the event.  We require
isolation $<0.1\times\Et^{\gamma}$ for $\Et^{gamma}<$~20~GeV, and
$<2.0$~GeV$+0.02\times(\Et^{\gamma} -20$~GeV) for $\Et^{\gamma}>$~20~GeV. 
In the
tracking chamber we require the scalar sum of the \ptt\ of all tracks
in a cone of 0.4 to be $<2.0$~GeV$+0.005\times\Et^{\gamma}$, where all values
of $\Et^{\gamma}$ are in GeV.}
e)~have a shower shape in the CES~\footnote{CES: Central EM Strip Chambers.} consistent with a single photon;
f)~have no other significant energy deposited nearby in the CES.

\begin{figure}[!h]
\centering
\resizebox{0.5\textwidth}{!}{%
\includegraphics*{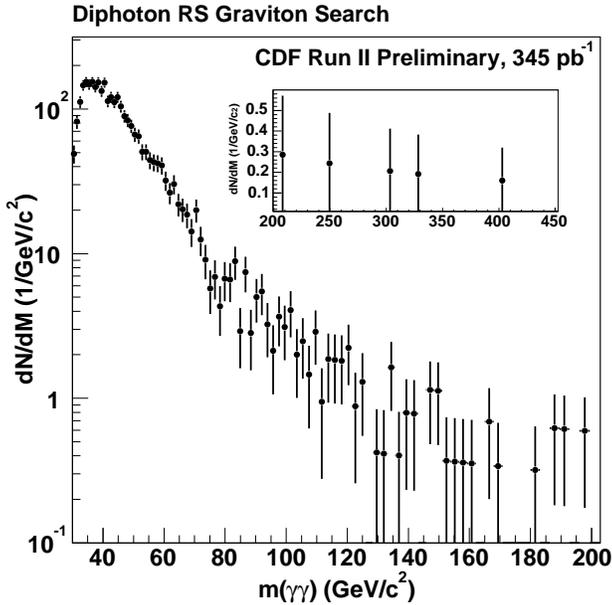}
}
\caption{The diphoton invariant mass distribution histogrammed in bins of
approximately one $\sigma$ of mass resolution.}
\label{mass_gg_varbin.figure}
\end{figure}

\begin{figure}[!h]
\centering
\resizebox{0.5\textwidth}{!}{%
\includegraphics*{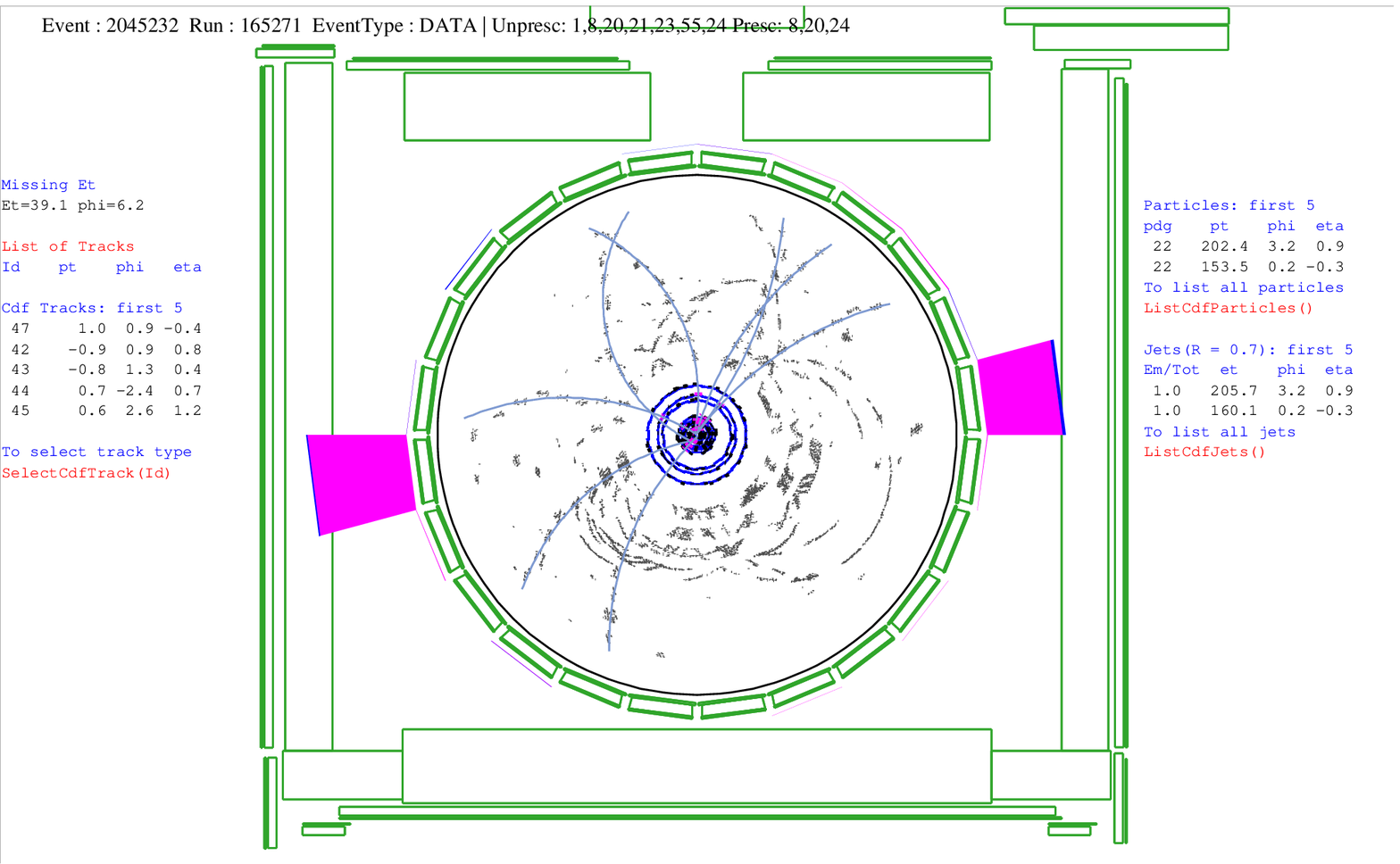}
}
\caption{
$\gg$ Highest Mass Event.
M ($\gg$)       = 405 GeV/$\rm{c^2}$,
$\Et^{\gamma1}$ = 172 GeV,
$\Et^{\gamma2}$ = 175 GeV.
}
\label{highest_mass.figure}
\end{figure}

The final dataset consists of 3339 events, for which the data
histogrammed with bins equivalent to one $\sigma$ of invariant mass
resolution are shown in Figure~\ref{mass_gg_varbin.figure}. The
highest mass events occur at masses of 207, 247, 304, 329, and 405
GeV/$\rm{c^2}$ (Figure~\ref{highest_mass.figure}).

\subsubsection{Backgrounds}

\begin{figure}[!h]
\centering
\resizebox{0.4\textwidth}{!}{%
\includegraphics*{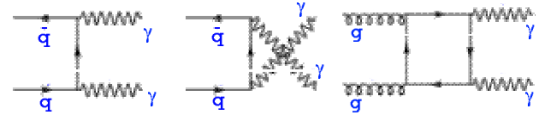}
}
\caption{Standard Model diphoton production diagrams.}
\label{gg_diagrams_sm.figure}
\end{figure}

There are two significant backgrounds to the $\gg$ sample. The first
is SM diphoton production which accounts for 30\% of the events
(Figure~\ref{gg_diagrams_sm.figure}). This background is estimated
using a NLO Monte Carlo, diphox~\cite{diphox}, which we normalize to
$\lum$=345 $\invpb$.

The second background comes from high-$\Et$ $\pi^{0}$'s from jets. To
create a control sample, we loosen several cuts (including relaxing
the isolation cuts by 50\%), and we get 9891 events, from which we
then reject events in the signal sample and are left with 6552 events
in the ``photon sideband'' sample. We then derive the shape of the
mass distribution by fitting this sample to a sum of several
exponentials. We then subtract the estimate from the SM contribution
and normalize the fakes background to the low mass ($m_{\gamma\gamma}$
between 30 and 100 GeV).

\begin{figure}[!h]
\centering
\resizebox{0.5\textwidth}{!}{%
\includegraphics*{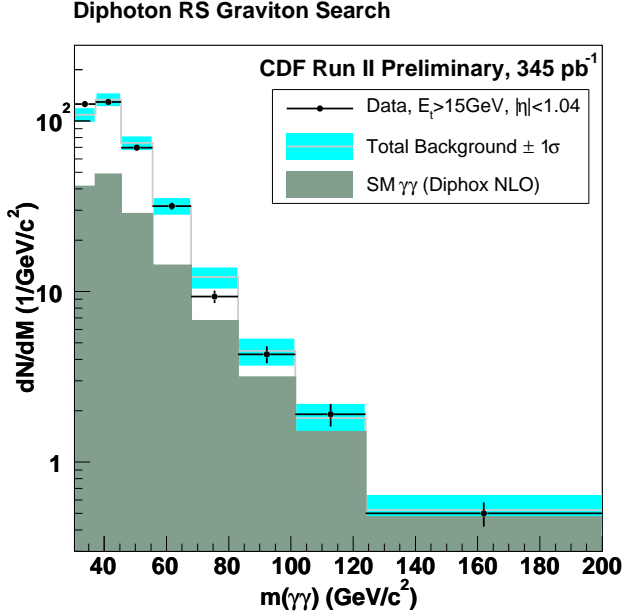}
}
\caption{
Comparison of the SM di-photon contribution plus misidentified jets
with the observed diphoton mass spectrum. Variable bins are used for
statistical comparison to the background prediction.  }
\label{mass_gg_bckgnd_log.figure}
\end{figure}

Figure~\ref{mass_gg_bckgnd_log.figure} shows the data mass spectrum
compared to the prediction.

\subsubsection{Limits on Randall-Sundrum Gravitons}

\begin{figure}[!h]
\centering
\resizebox{0.25\textwidth}{!}{%
\includegraphics*{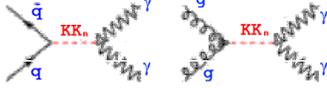}
}
\caption{Randall-Sundrum graviton production and decay to diphotons.}
\label{gg_diagrams_rs.figure}
\end{figure}

Since the data are consistent with the SM prediction, we place upper
limits on the cross sections times branching ratio of Randall-Sundrum
graviton production and decay to diphotons
(Figure~\ref{gg_diagrams_rs.figure}).

\begin{figure}[!h]
\centering
\resizebox{0.5\textwidth}{!}{%
\includegraphics*{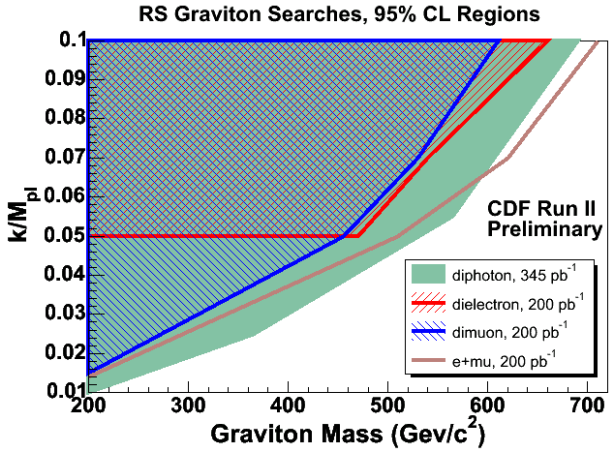}
}
\caption{Combined 95\% confidence level Randall-Sundrum graviton mass limits of the
di-photon and di-lepton searches.}
\label{comb_rs.figure}
\end{figure}

Figure~\ref{comb_rs.figure} shows the combined 95\% confidence level
RS graviton mass limits of the di-photon ($\lum$=345 $\invpb$) and
di-lepton ($\lum$=200 $\invpb$) searches~\cite{high_mass_dilepton} in
the graviton mass versus coupling, k/$M_{Planck}$, plane. Note, that
$\gg$ has a larger Branching Ratio (Br(G$\rightarrow\gg$) =
2$\times$Br(G$\rightarrow$ee)) and the $\gg$ spin factors improve
the acceptance.

\subsection{Search for Anomalous Production of Diphoton Events with $\met$ and Limits on GMSB Models}
\label{ggmet_runii.section}

For theoretical reasons~\cite{OurGMSB,SUSY_Workshop}, and because of
the $ee\gamma\gamma\met$ candidate event (Figure~\ref{eeggmet.figure})
recorded by the CDF detector in Run~I~\cite{Toback_PRD,Toback_PRL},
we want to search
for the production of heavy new particles that decay producing the
signature $\gamma\gamma+\met$.  Of particular theoretical interest
are supersymmetric (SUSY) models with gauge--mediated SUSY--breaking
(GMSB). Characteristically, the effective SUSY--breaking scale
($\Lambda$) can be as low as 100 TeV, the lightest SUSY particle is a
light gravitino ($\Gravitino$) that is assumed to be stable, and the
SUSY particles have masses in a range that may make them accessible at Tevatron
energies~\cite{OurGMSB}. In these models the visible signatures are
determined by the properties of the next--to--lightest SUSY particle
(NLSP) that may be, for example, a slepton or the lightest neutralino
($\none$).  In the GMSB model investigated here, the NLSP is a $\none$
decaying almost exclusively to a photon and a $\Gravitino$ that
penetrates the detector without interacting, producing $\met$. SUSY
particle production at the Tevatron is predicted to be dominated by
pairs of the lightest chargino ($\cone$) and by associated production
of a $\cone$ and the next--to--lightest neutralino ($\ntwo$). Each
gaugino pair cascades down to two $\NONE$'s, leading to a final state
of $\gamma\gamma+\met+X$, where $X$ represents any other final state
particles.

Details on this analysis can be found
in~\cite{cdf_ggmet,cdf_and_d0_ggmet_combined}.

\subsubsection{Data Sample}

The analysis selection criteria have been optimized to maximize, {\it
a priori}, the expected sensitivity to GMSB SUSY based only on the
background expectations and the predictions of the model.  Event
selection requirements for the diphoton candidate sample are designed
to reduce electron and jet/$\pi^0$ backgrounds while accepting
well-measured di\-photon candidates.

We require two central (approximately $0.05<|\eta|<1.0$)
electromagnetic clusters that should pass standard photon selection
cuts (Section~\ref{gg_runii_data.section}). For this analysis we
require $\Et^{\gamma}\gt$ 13 GeV.

\subsubsection{Backgrounds}
Backgrounds for the $\ggX$ analysis are:
\begin{itemize}
\item QCD background: fake photon (jj, j$\gamma$)
\item QCD background: $\gamma\gamma$
\item $e\gamma$
\item Non-Collision: beam-related, cosmic rays
\end{itemize}

Before the $\met$ requirement, the diphoton candidate sample is
dominated by QCD interactions producing combinations of photons and
jets faking photons. In each case only small measured $\met$ is
expected, due mostly to energy measurement resolution effects.

Events with an electron and a photon candidate ($W\gamma\rightarrow
e\nu\gamma$, $Wj\rightarrow e\nu\gamma_{fake}$, $Z\gamma\rightarrow
ee\gamma$, etc.) can contribute to the diphoton candidate sample when
the electron track is lost (by tracking inefficiency or
bremsstrahlung) to create a fake photon. For $W$ decays large $\met$
can come from the neutrinos. This background is estimated using
$e\gamma$ events from the data.

Beam--related sources and cosmic rays overlapped with a SM event can
contribute to the background by producing spurious energy deposits
that in turn affect the measured $\met$. While the rate at which these
events contribute to the diphoton candidate sample is low, most
contain large $\met$. The spurious clusters can pass photon cuts.

Backgrounds and observed number of events are summarized in
Table~\ref{ggmet.table}.

\begin{table*}[tbp]
\centering
\caption{Numbers of events observed and events expected from background sources
as a function of the \met requirement. Here ``QCD'' includes the
$\gamma\gamma$, $\gamma j$ and $jj$ processes.  The first uncertainty
is statistical, the second is systematic.
}
\label{ggmet.table}
\begin{tabular}{c|c|c|c|c||c}
\hline
\multicolumn{1}{c|}{\met }   & \multicolumn{4}{c||}{Expected} & Observed \\ \cline{2-5}
Requirement &  QCD & $e\gamma$ & Non-Collision & Total &  \\ 
\hline\hline
25 GeV  & $4.01\pm 3.21 \pm 3.76$ & $1.40 \pm 0.52 \pm 0.45$ & $0.54 \pm 0.06 \pm 0.42  $ & $5.95\pm 3.25 \pm 3.81$ & 3 \\ \hline
35 GeV  & $0.30\pm 0.24 \pm 0.22$ & $0.84 \pm 0.32 \pm 0.27$ & $0.25 \pm 0.04 \pm 0.19  $ & $1.39\pm 0.40 \pm 0.40$ & 2 \\ \hline
45 GeV  & $0.01\pm 0.01 \pm 0.01$ & $0.14 \pm 0.06 \pm 0.05$ & $0.12 \pm 0.03 \pm 0.09  $ & $0.27\pm 0.07 \pm 0.10$          & 0 \\ \hline
55 GeV  & (negligible)            & $0.05 \pm 0.03 \pm 0.02$ & $0.07 \pm 0.02 \pm 0.05  $ & $0.12\pm 0.04 \pm 0.05$         & 0 \\
\hline
\end{tabular}
\end{table*}          

\subsubsection{Limits on GMSB Models}

\begin{figure}[!h]
\resizebox{0.5\textwidth}{!}{%
\includegraphics*{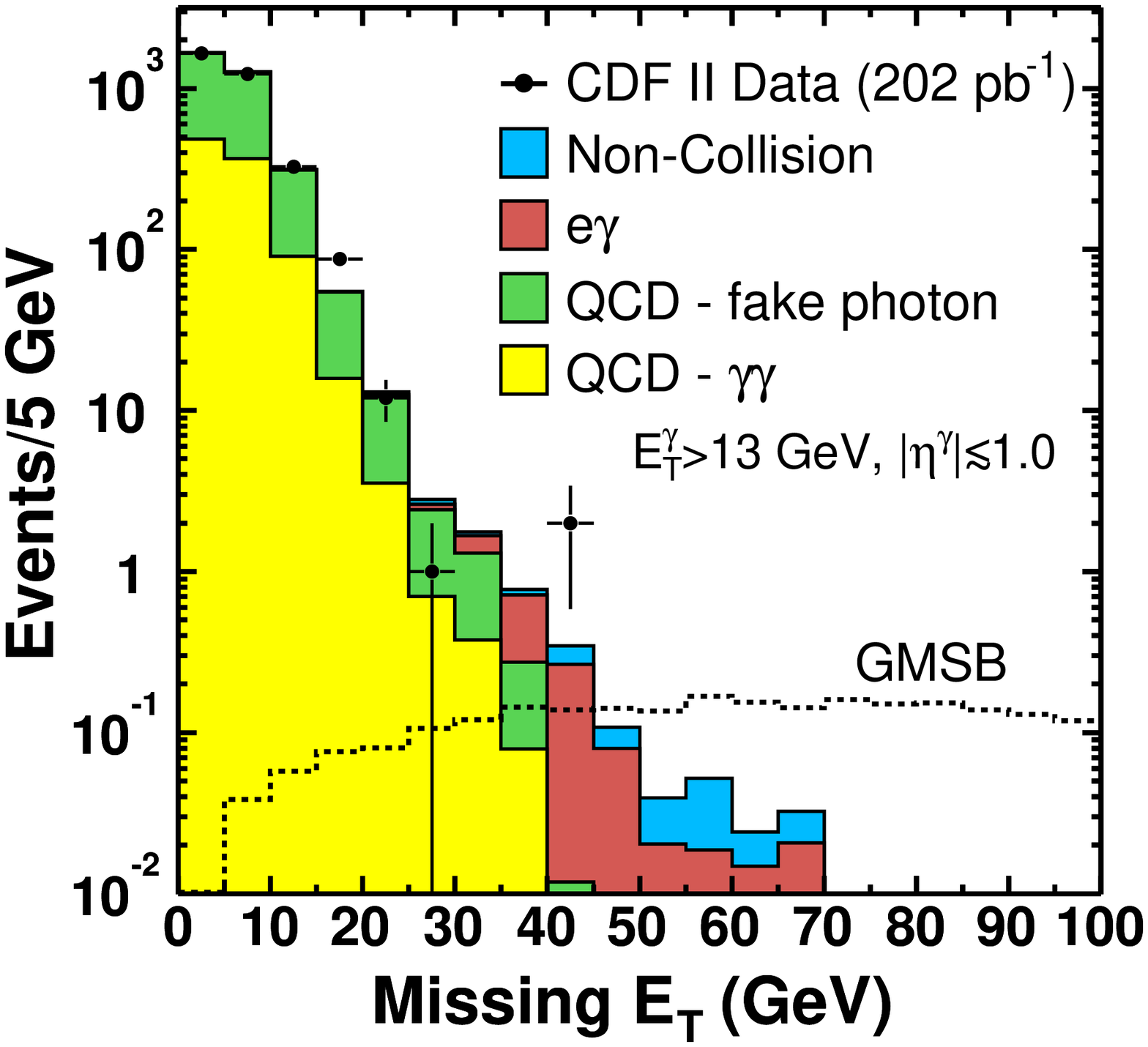}
}
\caption{
The $\met$ spectrum for events with two isolated central
photons with $\Et^{\gamma}>13$~GeV 
and \protect{\mbox{$|\eta|\lsim
1.0$}} along with the 
predictions from the GMSB model with a $\CONE$ mass of 175~GeV/$c^2$, normalized to 
202~$\invpb$.
The diphoton candidate sample data are in good agreement 
with the background predictions. There are no events above the 
$\met$ $>$~45~GeV threshold. The properties of the two candidates above 40 GeV appear consistent with the 
expected backgrounds.
}
\label{ggmet.figure}
\end{figure}

The $\met$ spectrum for events with two isolated central photons with
$\Et^{\gamma}>13$~GeV is shown in Figure~\ref{ggmet.figure}, along
with the predictions from the GMSB model. No excess is observed in two
photons + energy imbalance events.

Since there is no evidence for events with anomalous $\met$ in the
diphoton candidate sample, we set limits on new particle production
from GMSB using the parameters suggested in Ref.~\cite{snowmass}.
Using the NLO predictions we set a limit of M$_{\cone}$, and then from
mass relations in the model, we equivalently set limits on
M{$_{\none}$} and $\Lambda$:

M$_{\cone}\gt$167 GeV/$c^2$, M$_{\none}\gt$93 GeV/$c^2$, $\Lambda\gt$69 GeV/$c^2$.

The combined CDF+D\O\ limit~\cite{cdf_and_d0_ggmet_combined} is
significantly larger (i.e. more stringent) than either experiment
alone~\cite{cdf_ggmet,d0_ggmet}. The details on the combination of the
results on the CDF and D\O\ searches for chargino and neutralino
production in GMSB SUSY using the two-photon and missing $\Et$ channel
are explained in~\cite{cdf_and_d0_ggmet_combined}.

\begin{figure}[!h]
\centering
\resizebox{0.5\textwidth}{!}{%
\includegraphics*{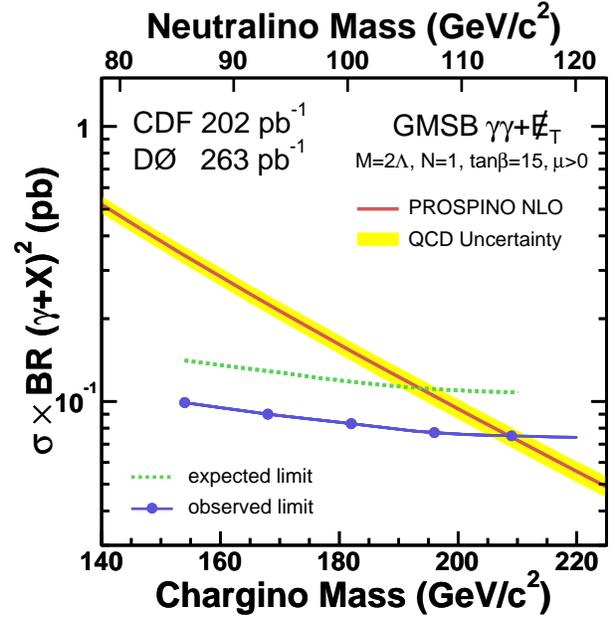}
}
\caption{The 95\% C.L. upper limits on the total production cross section 
times branching ratio versus M$_{\CONE}$ and M$_{\none}$ for the light
gravitino scenario using the parameters proposed
in~\protect\cite{snowmass}. The lines show the experimental combined
CDF+D\O\ limit and the LO and NLO theoretically predicted cross
sections. We set limits of M$_{\cone}\gt$209 GeV/$c^2$,
M$_{\none}\gt$114 GeV/$c^2$, $\Lambda \gt$84.6 GeV/$c^2$ at 95\%
C.L. in GMSB Model.}
\label{cdf_and_d0_limits.figure}
\end{figure}

Figure~\ref{cdf_and_d0_limits.figure} shows the combined CDF and D\O\
result for the observed cross section~\cite{cdf_and_d0_ggmet_combined}
as a function of M$_{\CONE}$ and M$_{\none}$ along with the
theoretical LO and NLO production cross sections.

The combined CDF+D\O\ limits are:\\
M$_{\cone}\gt$209$\GeVcsq$, M$_{\none}\gt$114$\GeVcsq$, $\Lambda\gt$84.6$\GeVcsq$ \\
at 95\% C.L. in GMSB Model. This is a first combined Run II result and
it sets the world's most stringent limits on the GMSB SUSY.

\subsection{Search for Lepton-Photon-X Events}
\label{lgx_runii.section}

In Run I lepton+photon+X search the results were consistent with
SM expectations in a number of channels with ``the
possible exception of photon-lepton events with large $\met$, for
which the observed total was 16 events and the SM
expectation was $7.6\pm0.7$ events, corresponding in likelihood to a
2.7 sigma effect.''~\cite{Jeff_PRL}). We concluded ``However, an
excess of events with 0.7\% likelihood (equivalent to 2.7 standard
deviations for a Gaussian distribution) in one subsample among the
five studied is an interesting result, but it is not a compelling
observation of new physics.  We look forward to more data in the
upcoming run of the Fermilab Tevatron.''~\cite{Jeff_PRL}. In this
section we report the results~\cite{lgx_runii} of
repeating the $\lgX$ search with the same kinematic selection
criteria in a substantially larger data set, $\lum$=305 $\invpb$,
a higher $\pbarp$ collision energy, 1.96 TeV, and the CDF II detector.

\subsubsection{Data Sample}
\label{lgx_data.section}

The data presented here were taken between March 21, 2002, and August
22, 2004 and represent 305$\invpb$ for which the silicon detector and
all three central muon systems (CMP, CMU and CMX) were operational.

A 3-level trigger~\cite{CDFII} system selects events with a high
transverse momentum~\footnote{We use the convention that ``momentum''
refers to $pc$ and ``mass'' to $mc^2$.} lepton ($\Pt > 18~\GeV$) or
photon ($\Et > 25~\GeV$) in the central region, $|\eta|
\lesssim 1.0$. Photon and electron candidates are chosen from clusters of
energy in adjacent CEM~\footnote{CEM: Central EM Calorimeter.} towers;
electrons are then further separated from photons by requiring the
presence of a COT~\footnote{COT: Central Outer Tracker.} track
pointing at the cluster. Muons are identified by requiring COT tracks
to extrapolate to a reconstructed track segment in the muon drift
chambers.

We have reused the Run I selection kinematic cuts for Run II analysis,
so that they are {\it a priori}:
\begin{itemize}
    \item {\it Tight} Muons: $\Pt>$ 25 GeV
    \item {\it Tight} Central Electrons, Photons: $\Et>$ 25 GeV
    \item {\it Loose} Muons: $\Pt>$ 20 GeV
    \item {\it Loose} Central Electrons: $\Et>$ 20 GeV
    \item {\it Loose} Plug Electrons: $\Et>$ 15 GeV
    \item $\met>$ 25 GeV
\end{itemize}

The identification of photons (see
Section~\ref{gg_runii_data.section}) and leptons is essentially the
same as in the Run I search~\cite{Jeff_PRD}, with only minor technical
differences, mostly due to the changes in the construction of the
tracking system and end-plug calorimeters.

A muon passing the `tight' cuts is required to: a) have a track in the
COT that passes quality cuts on the minimum number of hits on the
track; b) deposit energy in the electromagnetic and hadronic
compartments of the calorimeter consistent with that expected from a
muon, c) match a muon `stub' track in the CMX detector or in both the
CMU and CMP detectors; d) not be a cosmic ray (determined from
measuring timing with the COT).

`Tight' central electrons are required to have a high-quality track
with $\Pt$ of at least half the shower energy~\footnote{The $\Pt$
threshold is set to 25~$\GeV$ for $\Et > 100$ GeV.}, minimal leakage
into the hadronic calorimeter
\footnote{The fraction of electromagnetic energy $E_{em}$ allowed to leak into
the hadronic compartment is $0.055+0.00045E_{em}$ for tight and loose
central electrons; for loose plug electrons and for photons the
fraction must be less than 0.125.}, a good profile in the $z$
dimension (the dimension in which the electron track is not bent by
the magnetic field) at shower maximum that matches the extrapolated
track position, and a lateral sharing of energy in the two calorimeter
towers containing the electron shower consistent with that expected.

The additional muons are required to have $\Pt>20$ $\GeV$ and to
satisfy the same criteria as for ``tight'' muons but with fewer hits
required on the track, or, alternatively, a more stringent cut on
track quality but no requirement that there be a matching ``stub'' in
the muon systems. Additional central electrons are required to have
$\Et > 20~\GeV$ and to satisfy the tight central electron criteria but
with a track requirement of only $\Pt>10$ $\GeV$ (rather than
0.5$\times\Et$), and no requirement on a shower maximum measurement or
lateral energy sharing between calorimeter towers. `Loose' electrons
in the end-plug calori\-meters are required to have $\Et> 15$ GeV,
minimal leakage into the hadron calorimeters, a `track' containing at
least 3 hits in the silicon tracking system, and a shower transverse
shape consistent with that expected, with a centroid close to the
extrapolated position of the track.


Missing transverse energy $\met$ is calculated from the calorimeter
tower energies in the region $|\eta| < 3.6$. Corrections are then made
to the $\met$ for non-uniform calorimeter response~\cite{jet_corr} for
jets with uncorrected $\Et > 15$ $\GeV$ and $\eta < 2.0$, and for
muons with $\Pt > 20$ $\GeV$.

\subsubsection{Control Samples and Backgrounds}
We use $W$ and $Z^0$ production as control samples to ensure that the
efficiencies for high-$\Pt$ electrons and muons, as well as for
$\met$, are well understood. The photon control sample is constructed
from events in which one of the electrons radiates a high-$\Et$
$\gamma$ such that the $\eg$ invariant mass is within 10 $\GeV$ of the
$Z^0$ mass.

\begin{figure}[!t]
\centering
\resizebox{0.3\textwidth}{!}{%
\includegraphics*{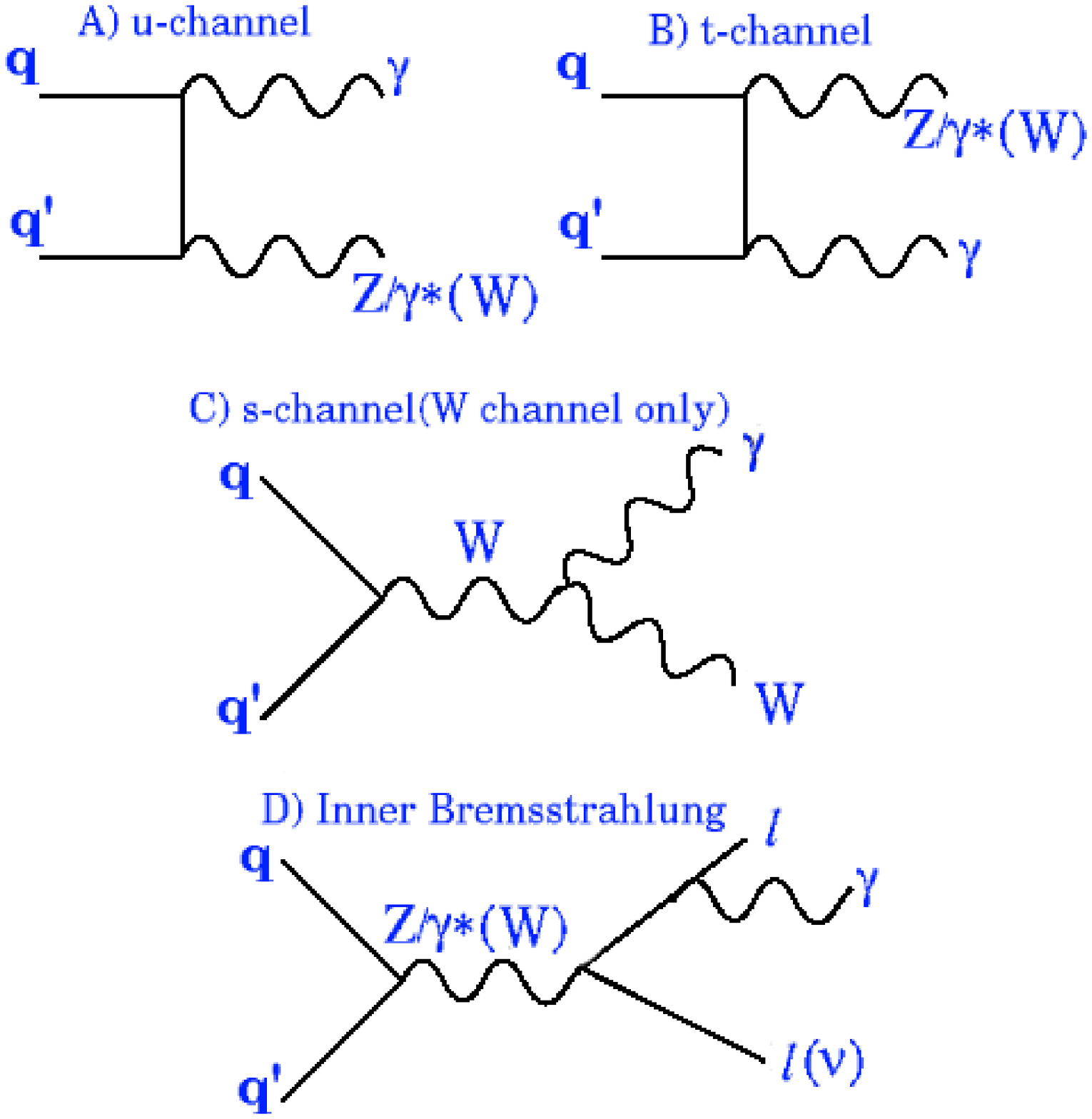}
}
\caption{Standard Model $\Wg$ and $\Zg$ production diagrams.}
\label{wg_zg_diagrams.figure}
\end{figure}

The dominant source of photon-lepton events at the Tevatron is
electroweak diboson production (Figure~\ref{wg_zg_diagrams.figure}),
in which a $W$ or $Z^0$ boson decays leptonically ($\ell \nu$ or
$\ell\ell$) and a photon is radiated from either an initial-state
quark, the $W$ or $Z^0$, or from a charged final-state lepton.  The
number of such events is estimated using leading-order (LO) matrix
element event generators~\cite{MadGraph,Baur,CompHep}. A correction
for higher-order processes (K-factor) has been
applied~\cite{Baur_NLO}.

To simulate the triboson channels $\Wgg$ and $\Zgg$ we have used
MadGraph~\cite{MadGraph} and CompHep\cite{CompHep}.

\subsubsection{Lepton-Photon-X Results}
\label{lgx_runii_results.section}

Following the Run I analysis strategy, we define the $\lgmet$
subsample by requiring that an event contains, in addition to the
central lepton and central photon, $\met> 25$ GeV. A second signal
subsample, the $\llg$ sample, is constructed by requiring, in addition
to the central lepton and central photon, a second `loose' lepton with
$\Et>25$ GeV. These two subsamples were selected as the search regions
of interest from the Run I results with the same kinematic selections;
these two searches in the Run II data are thus {\it a priori}. Both
sample selections are `inclusive', in that there are no requirements
on the presence or absence of other objects.

In addition to the expectations from real SM processes that produce
real lepton-photon events, there are backgrounds due to misidentified
leptons and photons, and also incorrectly calculated $\met$.

We consider two sources of fake photons: QCD jets in which a $\pizero$
or a photon from hadron decay mimics a direct photon, and electron
bremsstrahlung, in which an energetic photon is radiated off of an
electron which then has much lower energy and curls away from the
photon.

Backgrounds from fake leptons and/or fake missing $\Et$ ('QCD') we
estimate from a sample, in which we expect to have very little real
lepton content~\cite{Sacha_Kopp_thesis} by selecting on loose leptons
and rejecting events from the W or Z.

\begin{table}[!h]
\begin{center}
\caption{A comparison of the numbers of events predicted by the
SM and the observations for the $\lgmet$ and $\llg$ searches. The SM
predictions are dominated by $\Wg$ and $\Zg$
production~\cite{MadGraph,Baur,CompHep}. Other
contributions come from $\Wgg$ and $\Zgg$,
leptonic $\tau$ decays, and misidentified leptons, photons, or
$\met$.}
\label{summary_table} 
\begin{tabular}{l@{\extracolsep{0.0cm}}ccc}

\hline
\multicolumn{4}{c} {
{\bf Lepton+Photon+$\bf\met$ Events}}\\
\hline
{\bf SM Source} & {\bf   $e\gamma\met$} & {\bf   $\mu\gamma\met$} & {\bf   $(e+\mu)\gamma\met$} \\
\hline
$W^{\pm}\gamma$                         & 
13.70$\pm$1.89    &
 8.84$\pm$1.35    &
22.54$\pm$2.80    \\
$\Zgstar + \gamma$                              & 
1.16$\pm$0.40     &
4.49$\pm$0.64    &
5.65$\pm$1.03     \\
$W^{\pm}\gamma\gamma, \Zgstar \plus \gamma\gamma$   & 
0.14$\pm$0.02   &
0.18$\pm$0.02   &
0.32$\pm$0.03  \\
$W^{\pm}\gamma,\Zgstar\plus\gamma\goes\tau\gamma$     & 
0.71$\pm$0.18  &
0.26$\pm$0.08  &
0.97$\pm$0.22  \\
\hline
$W^{\pm}$+Jet faking $\gamma$                     & 
2.8$\pm$2.8       &
1.6$\pm$1.6       &
4.4$\pm$4.4       \\
$\Zgstar \goes \epem,e\goes\gamma$           & 
2.45$\pm$0.33    &
-                                       &
2.45$\pm$0.33                           \\
Jets faking $\ell+\met$                         & 
0.7$\pm$0.7   &
0.3$\pm$0.3   &
1.0$\pm$0.8   \\
\hline
{\bf Total} & 
{\bf 21.7$\pm$3.4}     & 
{\bf 15.7$\pm$2.2}     & 
{\bf 37.3$\pm$5.4}     \\

\hline
{\bf Observed} & 
{\bf   25}                          & 
{\bf   17}                          & 
{\bf   42}                          \\
\hline

\multicolumn{4}{c}{
{\bf Multi-Lepton+Photon Events}}\\ 
\hline
{\bf SM Source} & {\bf  $ee\gamma$} & {\bf  $\mu\mu\gamma$} & {\bf  $ll\gamma$}  \\
\hline
$\Zgstar + \gamma$            		          & 
12.50$\pm$1.53    &
 7.81$\pm$0.88    &
20.31$\pm$2.40    \\
$\Zgstar + \gamma\gamma$ 	     	& 
0.24$\pm$0.03  &
0.12$\pm$0.02 &
0.36$\pm$0.04  \\
\hline
$\Zgstar+$Jet faking $\gamma$   & 
0.3$\pm$0.3     &
0.2$\pm$0.2     &
0.5$\pm$0.5     \\
$\Zgstar \goes \epem,e\goes\gamma$           & 
0.23$\pm$0.09    &
-                                       &
0.23$\pm$0.09                           \\
Jets faking $\ell+\met$                         & 
0.6$\pm$0.6  &
1.0$\pm$1.0  &
1.6$\pm$1.2  \\
\hline
{\bf Total} & 
{\bf 13.9$\pm$1.7}    & 
{\bf  9.1$\pm$1.4}    & 
{\bf 23.0$\pm$2.7}   \\

\hline
{\bf Observed} & 
{\bf   19}                          & 
{\bf   12}                          &                 
{\bf   31}                          \\
\hline
\end{tabular}
\end{center}
\end{table}

The predicted and observed totals for both the $\lgmet$ and $\llg$
searches are shown in Table~\ref{summary_table}. We observe 42
$\lgmet$ events, versus the expectation of 37.3 $\pm$ 5.4 events. If
the Run I ratio of observed to expected, which was 16/7.6, had held
up, the 2.7 $\sigma$ excess observed in Run I would have resulted in
an observation of 79 $\pm$ 11 events when applying the same analysis
to the Run II data, versus the 42 events observed. In the $\llg$
channel, we observe 31 events, versus an expectation of 23.0 $\pm$ 2.7
events. No $e\mu\gamma$ events are observed.

\begin{figure}[!b]
\centering
\resizebox{0.5\textwidth}{!}{%
\includegraphics[angle=90]{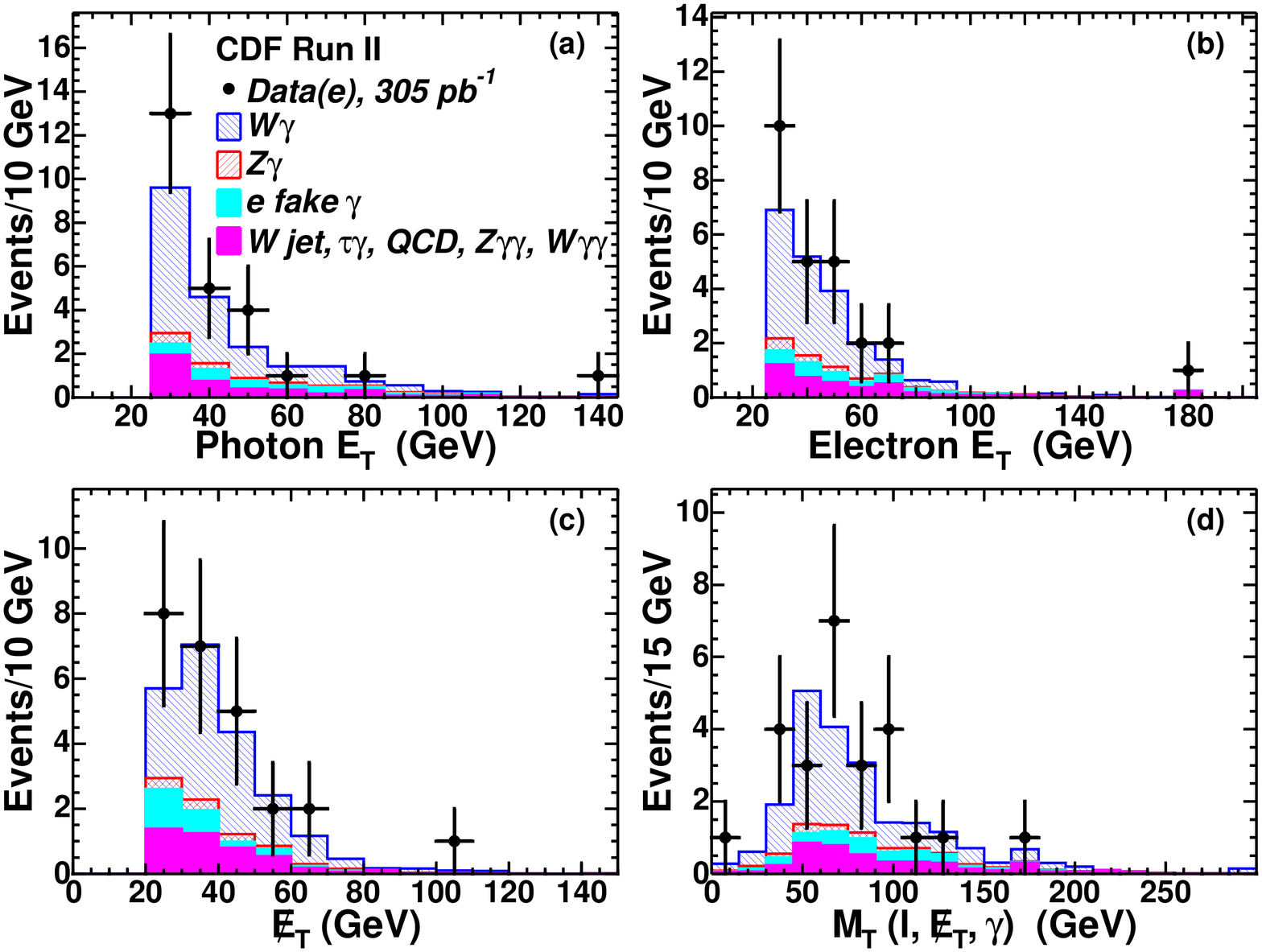}
}
\caption{
 The distributions for events in the $\egmet$ sample in a) the $\Et$
 of the photon; b) the $\Et$ of the electron, c) the missing
 transverse energy, $\met$, and d) the transverse mass of the
 electron-photon-$\met$ system. The histograms show the expected SM
 contributions, including estimated backgrounds from misidentified
 photons and leptons.}
\label{epj_figure_egmet.figure}
\end{figure}

While the number of events observed is somewhat larger than
expectations(Table~\ref{summary_table}), there is not a significant
excess in either signature, and the kinematic distributions are in
reasonable agreement with the SM predicted shapes.

\begin{figure}[!t]
\centering
\resizebox{0.5\textwidth}{!}{%
\includegraphics[angle=90]{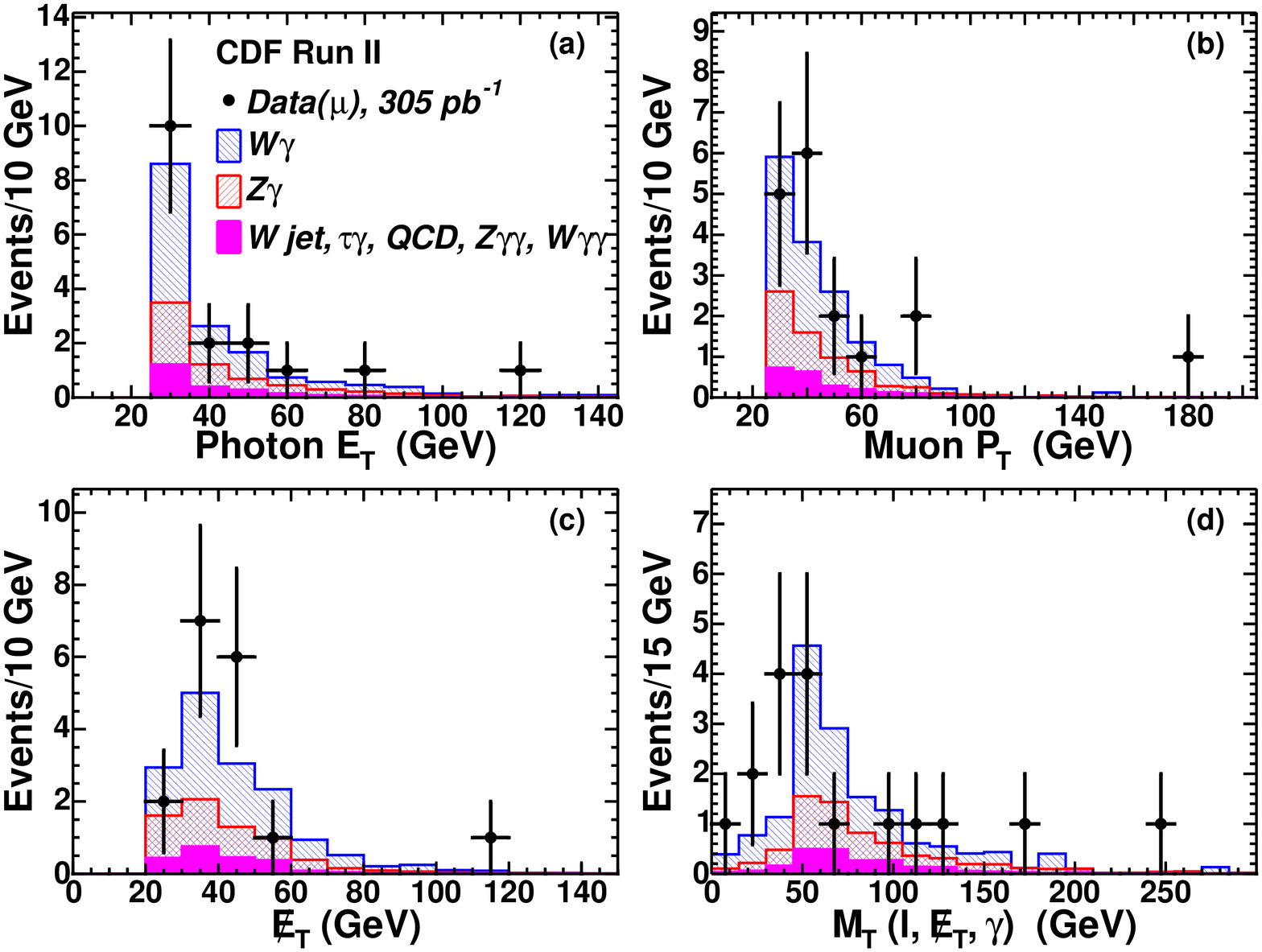}
}
\caption{
 The distributions for events in the $\mugmet$ sample in a) the $\Et$
 of the photon; b) the $\Pt$ of the muon, c) the missing transverse
 energy, $\met$, and d) the transverse mass of the muon-photon-$\met$
 system. The histograms show the expected SM contributions, including
 estimated backgrounds from misidentified photons and leptons.}
\label{epj_figure_mugmet.figure}
\end{figure}

The distributions for events in the $\lgmet$ sample are shown in
Figure~\ref{epj_figure_egmet.figure} for the electron channel and
in Figure~\ref{epj_figure_mugmet.figure} for the muon channel. The
dominant contribution for $\lgmet$ is SM $\Zg$ and $\Wg$
production.

\begin{figure}[!b]
\centering
\resizebox{0.5\textwidth}{!}{%
\includegraphics[angle=90]{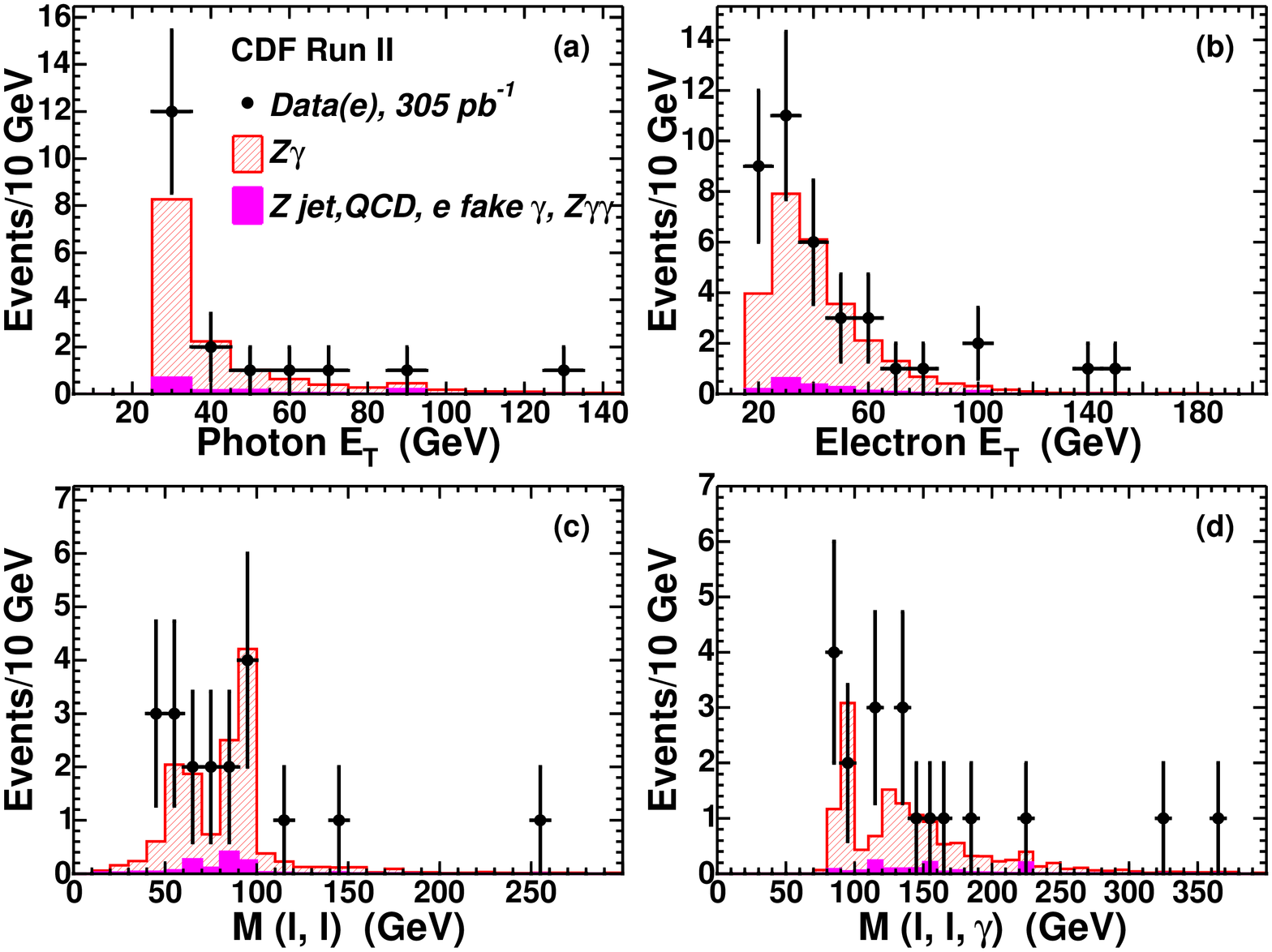}
}
\caption{
The distributions in a) the $\Et$ of the photon; b) the $\Et$ of the
electron, c) the 2-body mass of the dielectron system, and d) the
3-body invariant mass $m_{\eeg}$.}
\label{epj_figure_eeg.figure}
\end{figure}

\begin{figure}[!t]
\centering
\resizebox{0.5\textwidth}{!}{%
\includegraphics[angle=90]{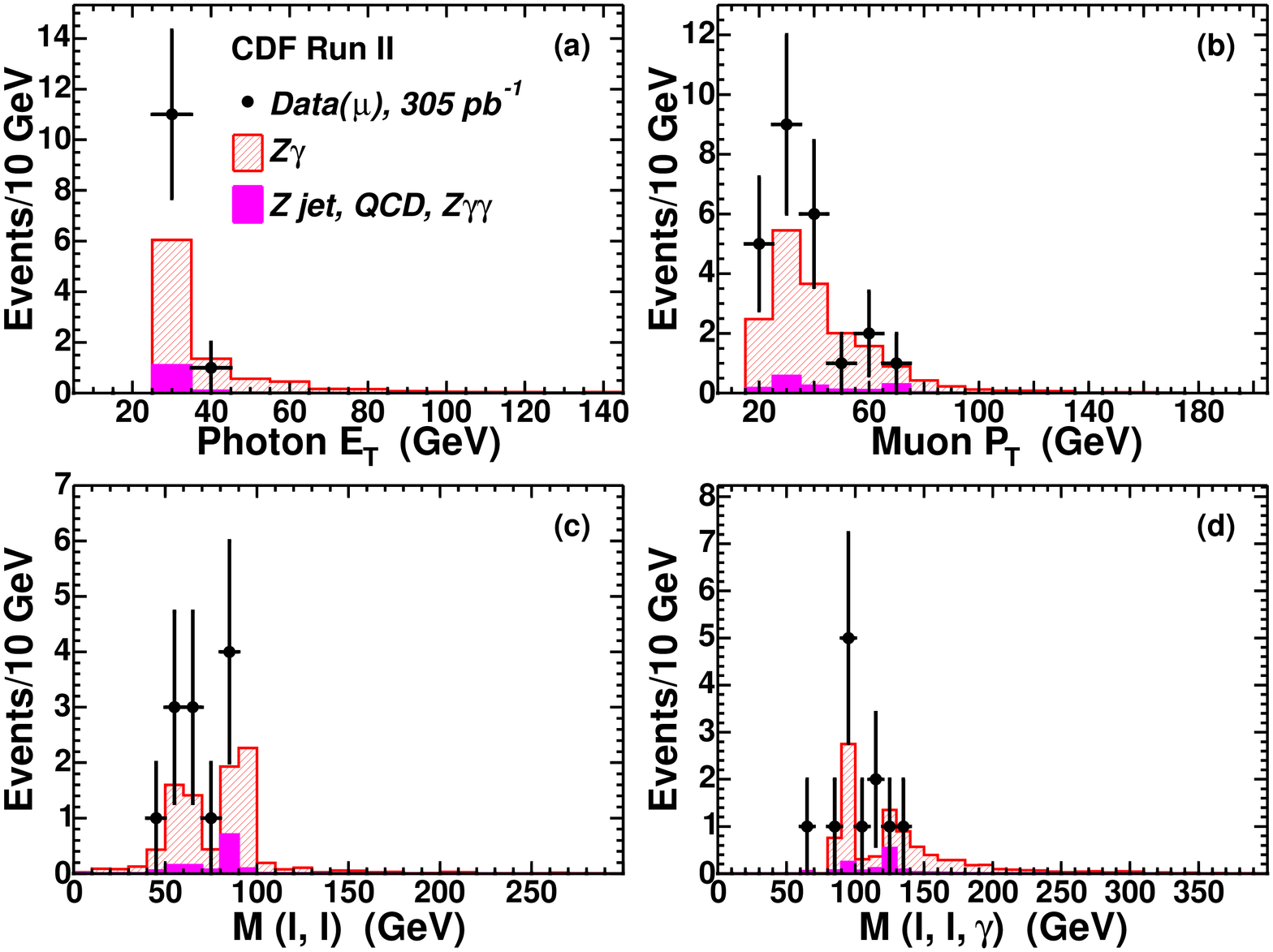}
}
\caption{
The distributions in a) the $\Et$ of the photon; b) the $\Pt$ of the
muon, c) the 2-body mass of the dimuon system, and d) the
3-body invariant mass $m_{\mumug}$.}
\label{epj_figure_mumug.figure}
\end{figure}

The distributions for events in the $\llg$ sample are shown at
Figure~\ref{epj_figure_eeg.figure} for electron channel and
Figure~\ref{epj_figure_mumug.figure} for muon channel. The dominant
contribution for $\llg$ is SM $\Zg$ production.

For the $\Zg$ process occurring via initial state radiation, the
dilepton invariant mass distribution will be peaked around the
$\Z$-pole. For the final state radiation, the three body invariant
mass (m(l, l, $\gamma$)) distribution will be peaked around the
$\Z$-pole (Figures~\ref{epj_figure_eeg.figure},~\ref{epj_figure_mumug.figure},
(c) and (d)).

\begin{figure}[!b]
\centering
\resizebox{0.5\textwidth}{!}{%
\includegraphics[angle=90]{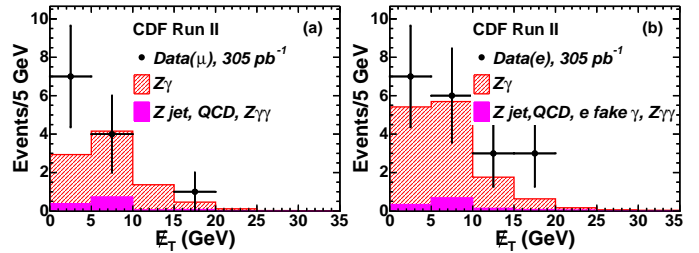}
}
\caption{The distributions in missing transverse energy $\met$
  observed in the inclusive search for a) $\mumug$ events and
  b) $\eeg$ events. The histograms show the expected SM
  contributions.}
\label{epj_figure_llg_met.figure}
\end{figure}

We do not expect missing $\Et$ in the events in the $\llg$ sample
based on the SM backgrounds; the $\eeggmet$ event was of special
interest due to the large value of
$\met$. Figure~\ref{epj_figure_llg_met.figure} shows the distributions
in $\met$ for the $\eeg$ and $\mumug$ subsamples of the $\llg$
sample. No events are observed with $\met> 25$ GeV.

In conclusion, we have repeated the search for inclusive lepton +
photon production with the same kinematic requirements as the Run I
search, but with a significantly larger data sample and a higher
collision energy. We find that the numbers of events in the $\lgmet$
and $\llg$ subsamples of the $\lgX$ sample agree with SM
predictions. We observe no $\llg$ events with anomalous large $\met$
or with multiple photons and so find no events like the $\eeggmet$
event of Run I.

In summary, while we are disappointed that we found no more $\eeggmet$
events in a much larger sample than in Run I, and the Run I excess in
$\lgmet$ became less significant rather than more, we have
conclusively settled a question that generated much interest in the
theoretical community. The channels we have investigated will remain
interesting, and the techniques we have developed and the knowledge
gained will be useful for similar searches at the Tevatron and at the
LHC.

\section{Summary and Outlook}
\label{summary_and_outlook.section}

To summarize, we will list the main points for the Run II results
presented in this paper:

\begin{itemize}
\item Search for $\lgX$: the Run I 2.7 sigma excess in $\lgmet$ is not
confirmed when repeating the analysis with much more data. We observe
no $\llg$ events with anomalous large $\met$ or with multiple photons.
\item Search for $\gg\met+X$: no excess is observed in the 
two photons + energy imbalance channel. The combined CDF and D\O\ Result provides world's most stringent limits on GMSB SUSY. No new $\eeggmet$ (or similar) candidate events have been found.
\item Search for high-mass diphotons: the data agree with predictions.
\end{itemize}

The Fermilab plan is to have a factor of 10-20 more data than
presented here by the end of Run II of the Tevatron. A recent upgrade,
the EM Timing system~\cite{em_timing_nim}, provides a vitally
important handle that could confirm (or disprove) that all the photons
in unusual events are from the primary collision.

Currently, the CDF is actively pursuing topics and analyzing up to 1
$\invfb$ of delivered luminosity. New and exciting results are coming
out quickly. Further information regarding the analyses presented in
this paper and new results can be found in~\cite{www_cdf_exotics}.

\end{document}